\documentclass[%
 reprint,
 amsmath,amssymb,
 aps,
]{revtex4-1}

\usepackage{graphicx}
\usepackage{dcolumn}
\usepackage{bm}

\usepackage{xcolor}
\usepackage{amsthm,mathtools}
\usepackage{upgreek}	

\newcommand{\R}{\mathbb{R}}
\newcommand{\C}{\mathbb{C}}
\newcommand{\N}{\mathbb{N}}
\renewcommand{\d}{\mathrm{d}}	

\let\Re\relax
\let\Im\relax
\DeclareMathOperator{\Re}{Re}
\DeclareMathOperator{\Im}{Im}

\newtheoremstyle{theorem}{}{}{\itshape}{}
{\bfseries}{}{\newline}{}

\theoremstyle{theorem}

\newtheoremstyle{example}{}{}{}{}
{\bfseries}{}{\newline}{}

\theoremstyle{definition}

\begin{document}


\title{Holomorphic Hamiltonian $\xi$-Flow and Riemann Zeros}

\author{Dirk~Lebiedz}
\affiliation{%
Institute for Numerical Mathematics \\
Ulm University, Germany
}%

\date{\today}

\begin{abstract}
\noindent
With a view on the formal analogy between Riemann-von-Mangoldts explicit formula and semiclassical quantum mechanics in terms of the Gutzwiller trace formula we construct a complex-valued Hamiltonian $H(q,p)=\xi(q)p$ from the holomorphic flow $\dot{q}=\xi(q)$ and its variational differential equation. The Hamiltonian phase portrait $q(p)$ is a Riemann surface equivalent to reparameterized $\xi$-Newton flow solutions in complex-time, its flow map differential is determined by all Riemann zeros and reminiscent of a 'spectral sum' in trace formulas. Canonical quantization for particle quantum mechanics on a circle leads to a Dirac-type momentum operator with discrete spectrum given by classical closed orbit periods determined by derivatives $\xi'(\rho_n)$ at Riemann zeros.
\end{abstract}

\maketitle


\section{Introduction}
\noindent
The idea to relate the nontrivial zeros of Riemanns $\zeta$-function to an eigenvalue problem of a (possibly self-adjoined) operator on some Hilbert space became famous as 'Hilbert-P\'{o}lya'-conjecture \cite{Polya1982}. With a view on a classical analogon of this operator, the 'Riemann dynamics', in particular the analogy \cite{Berry1999a} between Riemann-von-Mangoldt explicit formula for the Chebyshev version of the prime counting function and semiclassical quantum mechanics in terms of the Gutzwiller trace formula (see \cite{Gutzwiller1990}) renewed interest in a spectral interpretation of the Riemann zeros in the 1980s and 90s. Physical ideas like the Berry-Keating Hamiltonian and also more abstract mathematical approaches, e.g. by Connes \cite{Connes1996} and Deninger \cite{Deninger1998}, pointed out striking analogies to dynamical trace formulas. \\
Here we consider a classical mechanics viewpoint proposing a Hamiltonian system with a dense set of periodic orbits based on the holomorphic flow generated by $\xi(s)\in \C$, a complex 1-dimensional, real 2-dimensional vector field $(\Re(\xi(s),\Im(\xi(s))\in \R^2$ and its variational differential equation describing the differential of the $\xi$-flow map. Our Hamiltonian $H=\xi(q)p$ is a function of complex space and momentum variables $(q,p)\in \C^2$. Except for 'locally conformal scaling' of the space-variable $q$ by the entire function $\xi(q)$ it is the $\C$-analogon of the Berry-Keating Hamiltonian $H=xp, (x,p) \in \R^2$ \cite{Berry1999b,Sierra2011}.\\
We have been inspired by the qualitative and quantitative geometric results of phase portrait studies of the holomorphic $\zeta$- and $\xi$-flow \cite{Broughan2004,Broughan2005} as well as the  corresponding Newton-flows \cite{Neuberger2014,Neuberger2015,Schleich2018}. Here, our aim is to provide a setting in which both holomorphic $\xi$-flow and $\xi$-Newton flow in complex time and its Riemann surface solution manifold can be related to an appropriate Hamiltonian system, the Riemann zeros, prime numbers, the trace formula concept  in classical dynamical systems (see e.g. \cite{Cvitanovic1999}) and the corresponding number-theoretical analogon in terms of the Riemann-von-Mangoldt explicit formula \cite{Berry1999a}. Finally, we want to motivate further studies on the topology and geometry of the $\xi$-flow phase portrait by pointing out how it specifically encodes and comprises various properties of the $\xi$-function.

\section{$\zeta$-Newton flow in complex time}
\noindent
For $s \in \C$ and the logarithmic derivative of Riemanns $\zeta$-function in Euler-product representation of the Dirichlet series we find 
\begin{equation}\label{zetalogder}
 \frac{\zeta'(s)}{\zeta(s)} = -\sum_{m\in\N,p \text{ prime}} \ln p \cdot p^{-ms}, s\in \C, \Re(s) > 1. 
\end{equation}
Its backwards Newton flow equation in complex time $t\in \mathbb{C}$ (see \cite{Heitel2020} for motivation), which has the Riemann zeros as fixed-points, is
\[ \dot{s}=\frac{\d s}{\d t} =  \frac{\zeta(s)}{\zeta'(s)}. \]
It follows for $\Re(s) > 1$ 
\begin{equation*}
 \frac{\d t}{\d s} =  \frac{\zeta'(s)}{\zeta(s)} = -\sum_{m,p} \ln p \cdot p^{-ms} =  -\sum_{m,p} \ln p \cdot e^{-ms \ln p}. 
 \end{equation*}
The imaginary part of this Fourier-type sum is related to vibration frequences. Separation of variables and integration for $s= \sigma + i \tau=\Re(s) + i \Im(s) \in \C$ yields the solution 
\begin{eqnarray} 
t(s) &= &\sum_{m,p}  \frac{1}{m}  e^{-ms \ln p} \nonumber \\
&=&  \sum_{m,p} \frac{1}{m}  e^{-m \sigma \ln p} \cdot [ \cos(m \ln p \cdot \tau) \nonumber - i \sin (m \ln p \cdot \tau). ] 
\end{eqnarray}
Ignoring divergence of the $\zeta$-Dirichlet series on the critical axis  ($\sigma=\frac{1}{2}$) we get formally
\begin{equation}\label{fluctuations}
\Im (t(s)) = - \sum_{m,p} \frac{1}{m}  e^{-m \sigma \ln p}  \sin (m \ln p \cdot \tau). 
\end{equation}
\noindent
This is an infinite sum of modes whose amplitudes depend on the real part $\sigma$ and frequences determined by the imaginary part $\tau$ of $s \in\C$. Except for a constant scaling factor $\pi^{-1}$ eq. (\ref{fluctuations}) is exactly eq. (2.6) for the fluctuation term $\mathcal{N}_{fl}$ in Berry and Keating \cite{Berry1999a} for the couting function of Riemann zeros, when formally setting $\sigma=\frac{1}{2}$. So imaginary time in the complex-time $\zeta$-Newton flow solution seems to be directly related to prime numbers and the statistical distribution of Riemann-zeros. 

\subsection{$\xi$-Newton Flow}
\noindent
The Newton flows of both $\zeta$- and $\xi$-function have been studied in \cite{Neuberger2014,Neuberger2015,Schleich2018} providing some geometric insight into its phase space topology. Particularly this insight is based on 'lines of constant' phase, the complex argument (phase) of $\zeta(s)$ and $\xi(s)$ being a conserved property along Newton flow solution trajectories, and separatrices (lines separating the basins of attraction of distinct Riemann zeros). In \cite{Schleich2018} an equivalent formulation of the Riemann hypothesis is developed via a constant phase argument. \\
In \cite{Heitel2020} we introduced analytic continuation of the $\xi$-Newton flow to complex time in order to study Riemann solution surfaces instead of 1-dimensional trajectories which offer a considerably extended view on the properties of the dynamical system, in particular the global phase portrait topology. Here, we exploit the complex-time view to derive a formal relation between prime number logarithms and orbits of a classical dynamical system, an analogy that is central to the Gutzwiller trace formula interpretation of the Riemann-von-Mangoldt formula by Berry and Keating \cite{Berry1999a}. Recently we exploited in a similar sense imaginary time Fourier spectra to study global spectral properties of dynamical systems with multiple time scale structure \cite{Dietrich2020}. \\
Differentiating the defining equation
\begin{equation*}
    \xi(s) := \frac{1}{2}s(s-1)\Gamma\left(\frac{s}{2}\right)\pi^{-s/2} \zeta(s), s \in \C
\end{equation*}
the inverted Newton flows of $\zeta$ und $\xi$ can be related via 
\[  \frac{\xi'(s)}{\xi(s)} = \frac{1}{s} + \frac{1}{s-1} - \frac{1}{2} \ln \pi + \frac{\Gamma'(\frac{s}{2})}{\Gamma(\frac{s}{2})} + \frac{\zeta'(s)}{\zeta(s)} \]
with the digamma function $\psi(s):=\frac{\Gamma'(s)}{\Gamma(s)}$ and its explicit formula 
\[ \psi(s) = -\gamma + \sum_{n=0}^\infty \left(\frac{1}{n + 1} - \frac{1}{n + s}\right), s \in \C,s \neq 0, -1, -2, ...  \]
where $\gamma$ is the Euler-Mascheroni constant. The $\xi$-Newton flow in complex time has been introduced and discussed in  \cite{Heitel2020}. In particular, an explicit formula for its solution manifold was derived on the basis of the Hadamard product formula
$\xi(z)=\xi(0) \prod\limits_{n}(1-\frac{z}{\rho_n})$. 
Ignoring divergence of (\ref{zetalogder}) for $\Re(s)\le1$ and using $e^{-T}=\prod_n \frac{s(T)-\rho_n}{s_0-\rho_n}$ with the Riemann zeros $\rho_n$ from § 5.1 in\cite{Heitel2020} for the solution $s(T)$ of the Newton flow equation $\dot{s}= - \frac{\xi(s)}{\xi'(s)},s(0)=s_0$ we get 

\begin{eqnarray}
e^{-T}=\ln \frac{s}{s_0} + \ln \frac{s-1}{s_0-1} - \frac{1}{2} \ln \pi (s-s_0) - \gamma (s-s_0)   \nonumber \\
+ \sum_{n,p}  \frac{1}{n} \left( e^{ns \ln p} - e^{ns_0 \ln p} \right) \nonumber \\
+ \sum_n \left( \frac{s-s_0}{2n+2}-\ln (n+\frac{s}{2}) + \ln (n+\frac{s_0}{2}) \right)
\end{eqnarray}
\noindent
Except for the exponential sum, this formula contains only elementary terms and suggests a relation between prime numbers and complex time in the $\xi$-Newton flow solution.

\section{Hamiltonian System}
\noindent
We propose a Hamiltonian system $q,p\in \C, H:\C^2 \rightarrow \C, H:=\xi(q)p, t \in \R$ which connects  the holomorphic $\xi$-flow with both its variational flow and Newton flow.  
\begin{eqnarray}\label{hamiltonsyst}
\dot{q}=\frac{\d q}{\d t}&=&\frac{\partial H}{\partial p}=\xi(q), \quad q(0)=q_0 \nonumber \\
\dot{p}=\frac{\d p}{\d t}&=&-\frac{\partial H}{\partial q}=-\xi'(q)p, \quad p(0)=p_0. 
\end{eqnarray}
Phase space formulation with momentum $p$ leads to 
\begin{equation}\label{phasespaceODE}
\frac{\d q}{\d p}=\frac{\dot{q}}{\dot{p}}=-\frac{\xi(q)}{\xi'(q)}\cdot\frac{1}{p} 
\Leftrightarrow
\frac{\xi'(q)}{\xi(q)} \d q =- \frac{1}{p} \d p.
\end{equation}
\noindent
Separation of variables and use of the logarithmic derivative $(\ln \xi(q))'=\frac{\xi'(q)}{\xi(q)}$ of the $\xi$-product formula 
\begin{equation}\label{xiprodformel}
\xi(q)=\xi(0) \prod\limits_{n}\left(1-\frac{q}{\rho_n}\right)=\xi(0)\prod\limits_{n}\left(\frac{\rho_n-q}{\rho_n}\right)
\end{equation} 
with the $\xi$-zeros $\rho_n\in \C$ yields 
\[ \ln \xi(q)=\ln \xi(0)+ \sum_{n\in\N}\ln \left(\rho_n-q\right)- \sum_{n\in\N}\ln \rho_n \]
and $(\ln \xi(q))'= \sum_{n\in\N} \frac{1}{q-\rho_n}$. For $q(0)=q_0, p(0)=p_0$ integration of the Hamiltonian system (\ref{phasespaceODE}) leads to
\begin{equation}
\label{xiapprox}
    \begin{array}{rrlr}
	& \frac{\xi'(q)}{\xi(q)} \d  q&=-\frac{1}{p} \d  p  &\\
	\Leftrightarrow  & \int\limits_{q_0}^{q}\frac{\xi'(s)}{\xi(s)} \d  s &= -\int\limits_{p_0}^{p} \frac{1}{s} \d  s & \\
	\Leftrightarrow & \int\limits_{q_0}^{q} \sum\limits_{n} \frac{1}{s-\rho_n}\d  q &= \ln \frac{p_0}{p} &\\
	\Leftrightarrow & \sum\limits_{n}\int\limits_{q_0}^{q} \frac{1}{s-\rho_n}\d  q &=  \ln \frac{p_0}{p}  &\\
	\Leftrightarrow & \sum\limits_{n}\ln(q-\rho_n) - \sum\limits_{n} \ln(q_0-\rho_n) &=  \ln \frac{p_0}{p} &\\
	\Leftrightarrow & \sum\limits_{n} \ln \frac{q-\rho_n}{q_0-\rho_n} &=  \ln \frac{p_0}{p} & \\
	\Leftrightarrow & \exp\left(\sum\limits_{n} \ln \frac{q-\rho_n}{q_0-\rho_n}\right) & =   \frac{p_0}{p} \quad \mathrm{(mod } \ 2\pi i)&\\
	\Leftrightarrow & \prod\limits_{n} \frac{q-\rho_n}{q_0-\rho_n} &= \frac{p_0}{p} \quad \mathrm{(mod } \ 2\pi i)& 
    \end{array}
\end{equation}

\subsection{Algebraic Geometry Viewpoint}
\noindent 
Based on (\ref{xiapprox}) we define a polynomial $P_m \in \C[q,p],m \in \N$
\begin{eqnarray}
P_m (q,p,q_0,p_0):= \nonumber \\ 
p \prod\limits_{n=-m,n\ne 0}^m (q-\rho_n) - p_0\prod\limits_{n=-m,n\ne 0}^m (q_0-\rho_n)
\end{eqnarray}
with the product from $-m$ to $m$ referring to $m$ zeros together with their $m$ complex conjugated $\xi$-zeros.
We propose to consider the polynomial $P_m$ in complex projective space, which is for fixed initial values $(q_0,p_0) \in \C^2$ the projective plane $P^2_{\C}$ and the zero set $P_m(q,p)=0$ describes a plane algebraic curve on the Riemann sphere. Due to convergence of the $\xi$-product formula (\ref{xiprodformel})  (see \cite{Edwards2001} chap. 2 and \cite{Hadamard1893,VonMangoldt1895}), for $m\rightarrow \infty$ this algebraic set converges towards to solution manifold of the Hamiltonian flow (\ref{hamiltonsyst}).\\
The analytic and algebraic study of this manifold and its topology and symmetries might provide further insight into the properties of the $\xi$-function and the role of the Riemann zeros and their location. The work of Schleich et al. points out a significant role of the asymptotics of particular $\xi$-Newton flow lines, the separatrices (solution trajectories with a finite inverval of existence, see \cite{Broughan2003a}), for the location of the Riemann zeros. In \cite{Heitel2020} we suggest Poincar\'{e} sphere compactification of the dynamical system in order to study separatrices 'at infinity' and prove a result for the relation between fixed points and separatrices. We propose that the algebraic geometry viewpoint suggested here offers tools to characterize the separatrices as algebraic curves with particular geometric properties.

\subsection{Relation to $\xi$-Newton Flow}
\noindent
For momentum $p=p_0 e^T, T\in \C$ this result is in full agreement with the results from \cite{Heitel2020} for the $\xi$-Newton flow in complex time. The solution of $\dot{p}= -\xi'(q)p$ is $p(t)=p_0e^{-\int_0^t \xi'(q(\tau)) \d \tau}=p_0e^{\xi(q_0)-\xi(q)}$, i.e. $T= \xi(q_0)-\xi(q) +2 \pi k i, t\in \mathbb{R}, T \in \C, k\in \mathbb{Z}$. $q(T)$ can then be interpreted as the solution of a Newton-flow equation in complex time:
\begin{eqnarray}
\frac{\d q}{\d T} &=& \frac{\d q}{\d p}  \cdot \frac{\d p}{\d T}  = -\frac{\xi(q)}{\xi^\prime(q)} \nonumber \\
\frac{\d  T}{\d t} &=& \frac{\d  T}{\d p} \cdot \frac{\d  p}{\d t} =  \frac{1}{p} \cdot (-\xi^{\prime}(q)p) = - \xi^{\prime}(q). 
\end{eqnarray}
With substitution and logarithmic derivative we get 
\begin{eqnarray}
 \int\limits_0^t \xi^\prime(q(\tau))\d \tau &=& \int\limits_{q_0}^q \frac{\xi^\prime(s)}{\xi(s)}\d s  =
\ln(\xi(q))-\ln(\xi(q_0)) \nonumber \\
&=& \ln \left(\frac{\xi(q)}{\xi(q_0)} \right),
\end{eqnarray}
and for the time $T = -\ln(\xi(q(t)) + \ln(\xi(q_0)) + 2\pi k i$. The function $T(t)$ solves the differential equation
\begin{eqnarray}
\frac{\d T}{\d t} &=& -\frac{\d }{\d t} \ln (\xi(q(t)) = - \frac{1}{\xi(q(t))}\xi^\prime(q((t))\xi(q(t)) \nonumber \\
&=&  - \xi^\prime(q(t)).
\end{eqnarray}
This is obviously a nonlinear (differential) reparameterization of time $t\in \R \rightarrow T\in \C$ via $\xi'$ along solutions of the $\xi$-flow.

\subsection{Hamiltonian Variational Equation and Stability}
\noindent
The propagation of perturbations of initial values $\Delta q_0, \Delta p_0$ along closed Hamiltonian orbits is described to first order by the variational differential equation
\begin{eqnarray}
 \frac{\d}{\d t} \Delta q&=&\xi'(q) \Delta q, \Delta q(0)=\Delta q_0 \label{vardgl1} \\
 \frac{\d}{\d t} \Delta p&=&-\xi''(q)p \Delta q - \xi'(q) \Delta p, \Delta p(0)=\Delta p_0 
\end{eqnarray} 
Over the field $\C$ eq. (\ref{vardgl1}) and the momentum equation of (\ref{hamiltonsyst}) are 1-dim. and can be explicitly solved to
\begin{equation}
\Delta q= \frac{\xi(q)}{\xi(q_0)}\Delta q_0, \quad p= \frac{\xi(q_0)}{\xi(q)}p_0, \quad p\Delta q=  p_0 \Delta q_0
\end{equation} 
and so the $\Delta p$ equation is
\begin{equation}
 \frac{\d}{\d t} \Delta p=-\xi''(q) \Delta q_0 p_0  -\xi'(q) \Delta p
\end{equation} 
Variation of constant $c(t)$ with ansatz $\Delta p :=c(t) \xi^{-1}(q)$ leads to the solution
\begin{equation}\label{vardglsol}
\Delta p =  \frac{p_0\Delta q_0(\xi'(q_0)-\xi'(q))+\xi(q_0)\Delta p_0}{\xi(q)}
\end{equation} 
which turns out to be a summation formula using the $\xi$ product representation (\ref{xiprodformel}) and its logarithmic derivative:\begin{equation}\label{traceformula}
\Delta p =  p_0 \Delta q_0 \left( \frac{\xi'(q_0)}{\xi(q)} - \sum\limits_{n} \frac{1}{q-\rho_n}\right) + \frac{\xi(q_0)}{\xi(q)} \Delta p_0,
\end{equation} 
The perturbation $\Delta p$ propagates with reference to the Riemann zeros. Eq. (\ref{traceformula}) is a sensitivity equation allowing local stability analysis for Hamiltonian orbits and contains a 'spectral summation' reminiscent of a classical trace formula \cite{Cvitanovic1999}. 
For the complex-valued flow map differential matrix $M\in\C^{2\times 2}$ describing the evolution $(\Delta q, \Delta p)^{\top}=M(\Delta q_0, \Delta p_0)^{\top}$  of an initial perturbation along a given Hamiltonian orbit $(q(t),p(t))\in\C^2$ we get 
\begin{equation}\label{diffpoincaremap}
 M=
  \left( {\begin{array}{cc}
    \frac{\xi(q)}{\xi(q_0)} &  0\\
  p_0\frac{\xi'(q_0)}{\xi(q)}- \sum\limits_{n} \frac{p_0}{q-\rho_n} & \frac{\xi(q_0)}{\xi(q)}  \\
  \end{array} } \right)
\end{equation}
with a coupling constant $k_{q,p}:= p_0\frac{\xi'(q_0)}{\xi(q)}-\sum\limits_{n} \frac{p_0}{q-\rho_n}$ between $q$- and $p$-space. 
$M$ takes a particularly simple form for initial values in the set $\{ q_0 \in \C:\xi'(q_0)=0\}$, which are on separatrices of the $\xi$-Newton flow \cite{Neuberger2014,Neuberger2015}. 

\subsection{Action as Periodic Orbit Transit Time}
\noindent
For the action along a periodic Hamiltonian orbit we compute $S_p(E)=\oint p \ \d q$ with constant flow-invariant Hamiltonian (energy) $E=H(p,q)=H(p_0,q_0)$:
\begin{equation}
S_p(E)= \oint p(t) \xi(q(t)) \d t=\oint H(q,p) \d t=H(q_0,p_0) \cdot t^*
\end{equation}
with closed orbit period $t^*$ of the $\xi$-flow (see \cite{Broughan2005}). The energy-time duality formula (2.11) $t^*_p=\frac{\partial S_p}{\partial E}$ from \cite{Berry1999a} gives in our case the period $t^*=t^*_p=\frac{2\pi i}{\xi'(\rho_n)}$ with $n\in \N$ depending on $q_0$ fixing the periodic orbit. If the $\xi$-zero is simple, it is a center-type fixed point of the dynamical system \cite{Broughan2003}.
On a fixed energy level set invariant under the Hamiltonian flow the action is quantized. The quantization is determined by the derivatives $\xi'(\rho_0)$ at the Riemann zeros $\rho_n$ defining by elementary complex calculus \cite{Broughan2003,Broughan2005} the closed orbit period which is invariant under homotopy of periodic orbits. Broughan and Barnett already pointed out a potential role of $\xi$-flow closed orbit periods for the Hilbert-P\'{o}lya approach and numerically investigated a linear-scaling law for the periods with increasing imaginary part of the critical zeros with an upper bound for the periods \cite{Broughan2005a} under the Gonek conjecture assuming a lower bound for the $\zeta$-derivative $\zeta'(\rho_n)$ evaluated at the Riemann zeros. 

%

\section{Canonical Quantization}
\noindent
Consider point-particle quantum mechanics on a circle and a Dirac-type operator $D=\xi(q) \frac{\hbar}{i} \frac{\d}{\d q}$ of first order on the $\C$-Hilbert space $\mathcal{H}:=L^2(S^1, \d t)$, with $S^1$ parameterizing closed orbits of the $\xi$-flow with a Riemann metric corresponding to the line element $\d t = \frac{\d q}{\xi(q)}$ defined by the $\xi$-flow.
This is the momentum operator after canonical quantization $q \mapsto q \cdot, p \mapsto \frac{h}{i} \frac{\d}{\d q}$ of the Hamiltonian $H(q,p)=\xi(q)p$.\\
 The related eigenvalue problem is
\begin{eqnarray}
D f &=& E f, f\in \mathcal{H}, E\in \C 
\Leftrightarrow \xi(q)
\frac{\hbar}{i} \frac{\d f(q)}{\d q} = E f(q) \nonumber \\ 
 \frac{\d f(q)}{\d q}&=&\frac{i}{\hbar} \frac{E}{\xi(q)}\cdot f(q) \nonumber 
 \end{eqnarray}
 with $f \in \mathcal{H}$, i.e. periodic boundary conditions (particle on a circle). It follows for $\dot{q}=\xi(q)$
\[ f(q(t^*))=f(q(0)) e^{\frac{i}{\hbar} E \oint \d t} = f(q_0)e^{\frac{i}{\hbar} Et^*} \] 
with period $t^*\in \R$ and transit time $t^* < \infty$ for a simple zero $\rho_n$, $f(q(t^*))=f(q(0))$ and \[ \frac{i}{\hbar} E t^* = 2\pi i k, k \in \mathbb{Z}
\Leftrightarrow E=\frac{hk}{t^*} = k h \nu, k\in \ \mathbb{Z} \]
with frequency $\nu:=\frac{1}{t^*}$.
\noindent
So the eigenvalues of the Dirac-type momentum operator $D$ on $\mathcal{H}$ with classical counterpart Hamiltonian $H(q,p)=\xi(q)p$ are determined by periods of closed orbits of the $\xi$-flow encircling a simple $\xi$-zero (see \cite{Broughan2005}) and defining a quantum oscillator.

\section*{Acknowledgement}
\noindent
The author thanks Mazen Ali, Marcus Heitel and Johannes Poppe for comments and discussions and the Klaus-Tschira foundation (project 00.003.2019) for financial support.

\bibliographystyle{unsrt3Names}
\bibliography{literature.bib}

\end{document}